\documentclass[sigconf]{acmart}
	
\usepackage{bm}
\usepackage{amsmath}

\makeatletter

\newcommand\xleftrightarrow[2][]{%
	\ext@arrow 9999{\longleftrightarrowfill@}{#1}{#2}}
\newcommand\longleftrightarrowfill@{%
	\arrowfill@\leftarrow\relbar\rightarrow}
\makeatother
\newcommand{\minitab}[2][l]{
	\begin{tabular}{#1}#2\end{tabular}
}

\usepackage{cases}
\usepackage{subfigure}
\usepackage{enumitem}
\setlist[itemize]{leftmargin=*}
\usepackage{multirow}
\usepackage{makecell}
\usepackage{amstext}
\AtBeginDocument{%
	}
		
\copyrightyear{2023}
\acmYear{2023}
\setcopyright{acmlicensed}\acmConference[SIGIR '23]{Proceedings of the 46th International ACM SIGIR Conference on Research and Development in Information Retrieval}{July 23--27, 2023}{Taipei, Taiwan}
\acmBooktitle{Proceedings of the 46th International ACM SIGIR Conference on Research and Development in Information Retrieval (SIGIR '23), July 23--27, 2023, Taipei, Taiwan}
\acmPrice{15.00}
\acmDOI{10.1145/3539618.3591720}
\acmISBN{978-1-4503-9408-6/23/07}

\begin{document}
\begin{sloppypar}
\title{M2GNN: Metapath and Multi-interest Aggregated Graph Neural Network for Tag-based Cross-domain Recommendation}


%
\author{Zepeng Huai}
\affiliation{%
	\institution{School of Artificial Intelligence, University of Chinese Academy of Sciences}
	\city{Beijing}
	\country{China}
}
\email{zepenghuai6@gmail.com}

\author{Yuji Yang}
\affiliation{%
	\institution{Meituan}
		\city{Beijing}
	\country{China}
}
\email{yangyujiyyj@gmail.com}

\author{Mengdi Zhang}
\affiliation{%
	\institution{Meituan}
		\city{Beijing}
	\country{China}
}
\email{zhangmengdi02@meituan.com}

\author{Zhongyi Zhang}
\affiliation{%
	\institution{Meituan}
		\city{Beijing}
	\country{China}
}
\email{zhangzhongyi02@meituan.com}

\author{Yichun Li}
\affiliation{%
	\institution{Meituan}
		\city{Beijing}
	\country{China}
}
\email{yichun.li@meituan.com}

\author{Wei Wu}
\affiliation{%
	\institution{Meituan}
		\city{Beijing}
	\country{China}
}
\email{wuwei19850318@gmail.com}

\begin{abstract}
	Cross-domain recommendation (CDR) is an effective way to alleviate the data sparsity problem. Content-based CDR is one of the most promising branches since most kinds of products can be described by a piece of text, especially when cold-start users or items have few interactions. However, two vital issues are still under-explored: (1) From the content modeling perspective, sufficient long-text descriptions are usually scarce in a real recommender system, more often the light-weight textual features, such as a few keywords or tags, are more accessible, which is improperly modeled by existing methods. (2) From the CDR perspective, not all inter-domain interests are helpful to infer intra-domain interests. Caused by domain-specific features, there are part of signals benefiting for recommendation in the source domain but harmful for that in the target domain. Therefore, how to distill useful interests is crucial. To tackle the above two problems, we propose a \textbf{m}etapath and \textbf{m}ulti-interest aggregated \textbf{g}raph \textbf{n}eural \textbf{n}etwork (\textit{M2GNN}). Specifically, to model the tag-based contents, we construct a heterogeneous information network to hold the semantic relatedness between users, items, and tags in all domains. The metapath schema is predefined according to domain-specific knowledge, with one metapath for one domain. User representations are learned by GNN with a hierarchical aggregation framework, where the intra-metapath aggregation firstly filters out trivial tags and the inter-metapath aggregation further filters out useless interests. Offline experiments and online A/B tests demonstrate that M2GNN achieves significant improvements over the state-of-the-art methods and current industrial recommender system in Dianping, respectively. Further analysis shows that M2GNN offers an interpretable recommendation.
\end{abstract}



\maketitle

\section{Introduction} \label{sec:introduction}
\begin{figure}[h]
	\centering
	\includegraphics[width=\linewidth]{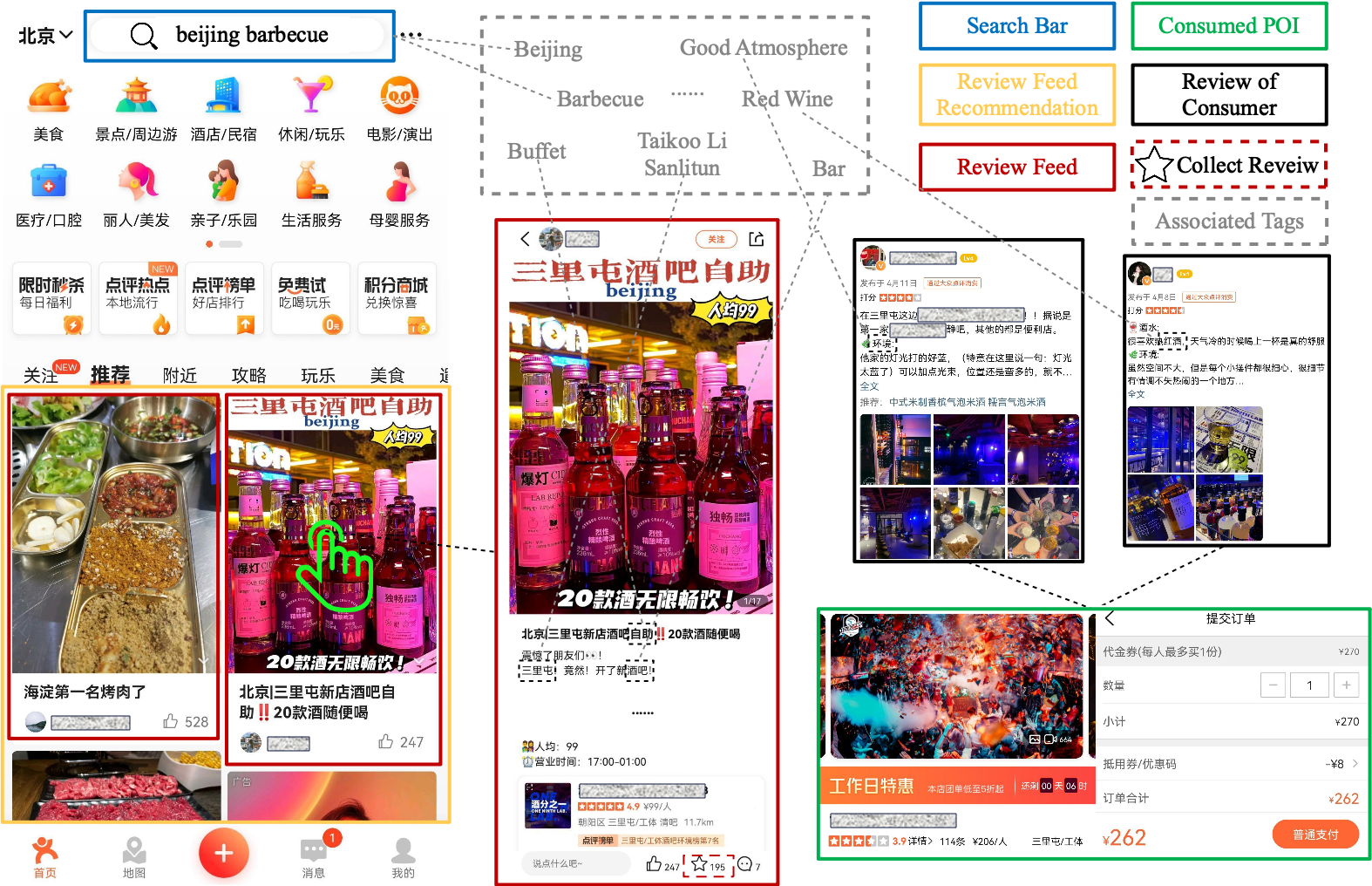}
	\caption{Dianping review feed recommendaiton and some examples to explain how tags are associated with the review, search, and consumption domain. Best viewed in color.}
	\Description{Dianping review feed recommendaiton.}
	\label{fig:dp_reveiw_feed}
\end{figure}
\begin{figure}[t]
	\centering
	\vspace{-0.35cm} 
	\subfigtopskip=0pt 
	\subfigbottomskip=0pt 
	\subfigcapskip=-5pt 
	\subfigure[\textbf{Data sparsity of collection log.}]{
		\includegraphics[width=0.5\linewidth]{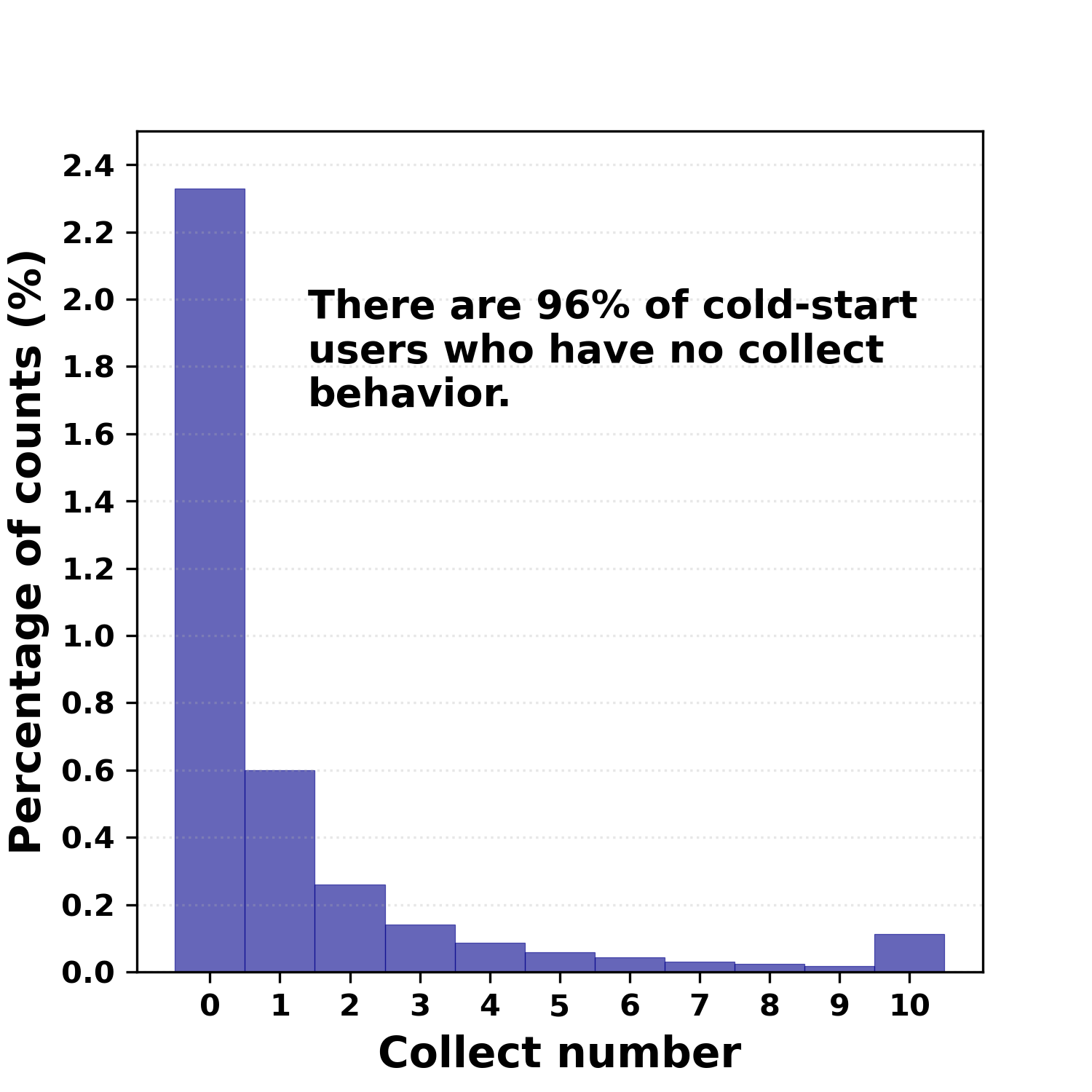}
	}\hspace{-4mm}
	\subfigure[\textbf{Tag distribution across the users in three domains.}]{
		\includegraphics[width=0.5\linewidth]{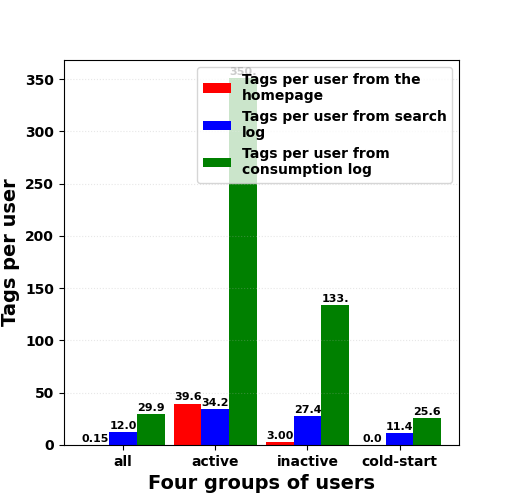}
	}\hspace{-4mm}
	\caption{User cross-domain behavior analysis.}
	\label{fig:Data feature}
\end{figure}

Dianping is one of the most popular online platforms in China\footnote{Monthly, hundreds of million UV and ten million user-generated reviews are accumulated.} for discovering local life services, including dining, shopping, entertainment, hotel, etc.. With the purpose of seeking trusted local business information, users are eager to search and consume in Dianping and write their reviews to share experiences. To increase user retention rates, Dianping establishes a review feed recommender system on the homepage as shown on the left side of Figure \ref{fig:dp_reveiw_feed}, which faces a severe data sparsity problem in the collection log\footnote{Here, our goal is to improve the collection rate rather than the click-through rate, because the former is the most important indicator for profit growth in our business data analysis.}. Take one-month data (Nov. 2022) as an example, Figure \ref{fig:Data feature}(a) shows the distribution of collection numbers across the users. Cold-start users account for about 96\% of all users, while active users (with more than ten collect histories) less than 0.2\%. 

To solve this dilemma, we introduce the more accessible light-weight textual side information called tags. Tag is a word or phrase, which refers to an object or activity, like hot pot and skiing. Take Figure \ref{fig:dp_reveiw_feed} as an example, we input \textit{beijing barbecue} in the search bar (blue box) and the associated tags are \textit{beijing} and \textit{barbecue}. Obviously, user tag-based interests extracted from other domains are useful for suggesting reviews on the homepage. For example, a user might get interested in a bar review because he searches for wines. 
We count tags associated with three domains (review, search, and consumption) via the same one-month data as shown in Figure \ref{fig:dp_reveiw_feed}. The results are in Figure \ref{fig:Data feature}(b) and show that users have rich behaviors in the latter two domains, especially for cold-start and inactive users. Therefore, \textbf{(Q1) how to leverage tag-based contents to enhance recommendations in the target domain is a key issue.} 

\begin{figure}[t]
	\centering
		\vspace{-0.35cm} 
	\subfigtopskip=0pt 
	\subfigbottomskip=0pt 
	\subfigcapskip=-5pt 
	\includegraphics[width=\linewidth]{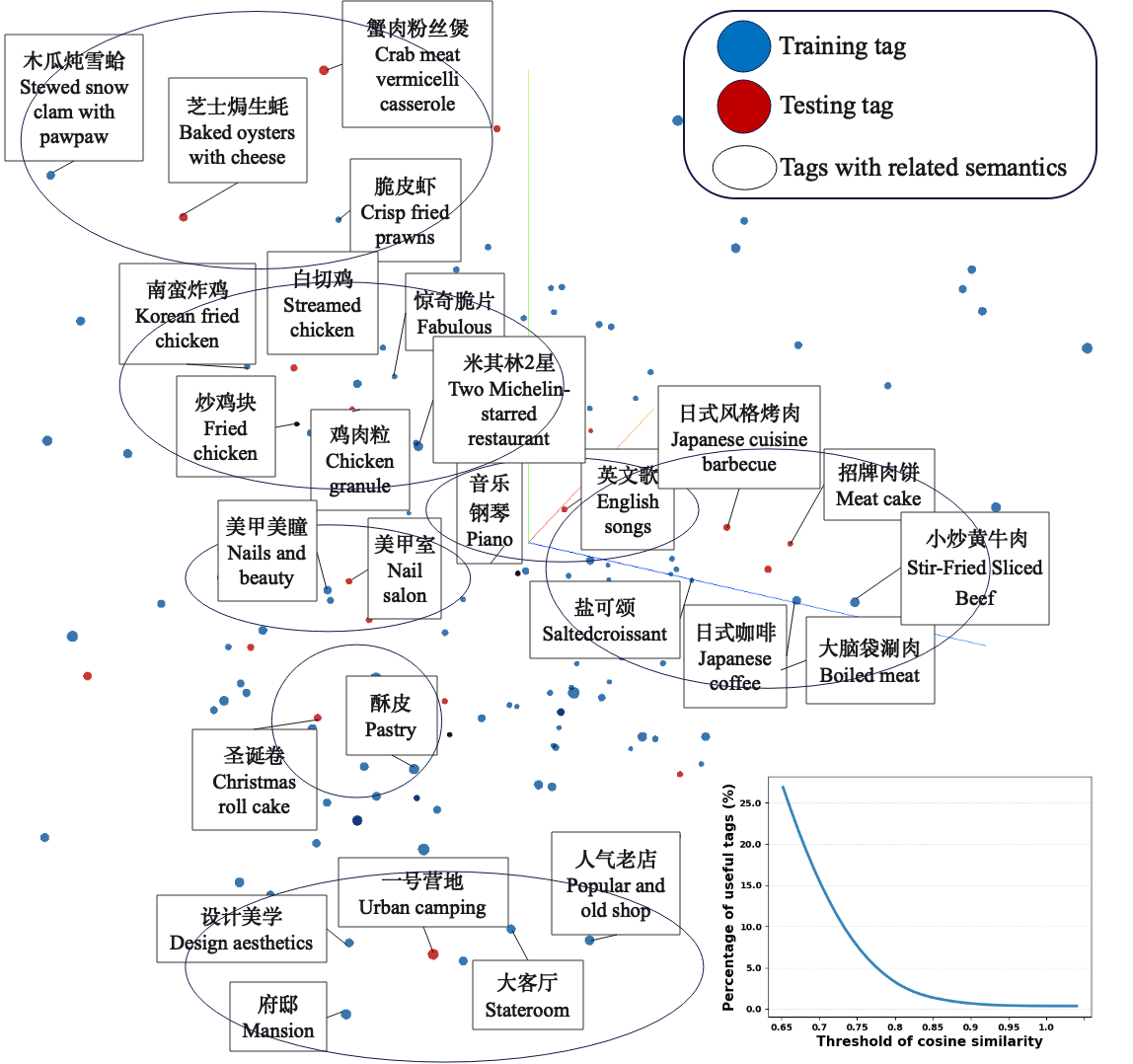}
	\caption{The proportion of useful tags and an example.}
	\Description{The proportion of useful tags and an example.}
	\label{fig:userful_tag}
\end{figure}

Furthermore, we analyze whether all associated tags of search and consumption are advantageous for review feed recommendation, that is, the proportion of useful tags. We gather tags associated with the above three domains in Nov. 2022, called training tags, and then collect tags just associated with reviews in the following week (from Dec. 1, 2022 to Dec. 7, 2022), called testing tags. All tags are embedded by SBERT \cite{reimers2019sentence}, which is pretrained on Dianping corpus specifically. For testing tags, we mine the relevant training tags by setting a threshold of cosine similarity between two embeddings. Figure \ref{fig:userful_tag} visualizes tag embeddings of a randomly selected user and shows the percentage of useful tags by averaging all users at the bottom right, which reveals a reasonable phenomenon that only a small part of tags (less than 10\% when the threshold is 0.7) can represent user interests towards reviews. Therefore, \textbf{(Q2) another pivotal problem is to distill a fraction of useful tags representing intra-domain interests from a large quantity, most of which are noises.}

Cross-domain recommendation (CDR) is an intuitive branch to solve the above problems. Early research \cite{singh2008relational,yan2019deepapf} focus on different mapping strategies for user cross-domain preferences. However, all of them don't introduce content information. To achieve content enhanced recommendation, some single-domain workings realize great success \cite{seo2017interpretable,wu2019context,zheng2017joint,xia2019leveraging}. Based on the previous efforts, content enhanced CDR methods firstly extract content-aware interests and then transfer them across domains. CATN \cite{zhao2020catn} adopts text convolution to learn aspect-level features, which are matched between the source and target domain via a correlation matrix. MTNet \cite{huang2016mtnet} attentively distills useful contents via a memory network and selectively feeds them into a transfer network. However, these approaches neglect how to model tag-based contents. In recent years, graph neural network (GNN) based CDR has attracted considerable research attention for its superiority in modeling high-order connections, like PPGN \cite{zhao2019cross} and HecRec \cite{yin2019heterogenous}. However, the aggregation via the GNN layer in the above methods overlooks noisy neighbor nodes, which are indeed harmful when the unuseful nodes account for the most.

In this paper, we propose a metapath and multi-interest aggregated graph neural network for tag-based cross-domain recommendation, called M2GNN. To solve \textbf{Q1}, we leverage the ability of graph to recognize the pattern of non-euclidean data \cite{zhou2020graph} including tag-based contents. Technically, a tag-associated heterogeneous graph is established to model the multi-homed behaviors, where each domain is represented by a metapath. Then a GNN-based method is designed to explore high-order tag-based cross-domain interests for enhancing recommendation in the target domain. To address \textbf{Q2}, we introduce a novel two-step aggregation mechanism employing the dynamic routing network and self-attention network. The former filters out the trivial tags and learns multiple high-level user interests in each domain. The latter furthermore transfers more significant interests into the target domain. Moreover, we introduce the skip-gram regularization to model the semantic correlation between tags. To summarize, the main contributions of this work are as follows:
\begin{itemize}
	\item We propose a novel CDR problem whose side information is tags and design a GNN-based method to solve it.
	\item We present a hierarchical aggregation schema to learn user preferences when noisy neighbor nodes are in the majority.
	\item We conduct offline experiments on a newly proposed industrial dataset and a public dataset, which proves the effectiveness and interpretability of M2GNN. Online A/B tests demonstrate the improvement to the current industrial recommender system.
\end{itemize}

\section{Preliminaries}
\subsection{Problem Setup} \label{sec:problemsetup}
\textbf{Tag-based cross-domain recommendation (TCDR)}. We have a target domain (e.g. review collection) and some source domains (e.g. search and consumption), denoted as $D^t$ and $D^s$, where $S=\{s_1,s_2,...\}$ represents there are $|S|$ source domains. Each domain includes its users $u$, items $i$, and interactions $\hat{y}_{ui}$. Note that users are full overlap, also known as multi-homed users, while items are non-overlap. We identify the item in different domains by its superscript, such as $i^t$, $i^{s_1}$, and $i^{s_2}$. 
According to interaction numbers in the target domain, users are divided into three groups, which are cold-start, inactive and active users. Let $U^{cs}$, $U^{ia}$ and $U^{a}$ represent them, respectively. In this paper, the bound to distinguish the latter two groups is set to be 10 times. Cold-start users have no interactions with items in $D^t$ , but with items in $D^s$. Each item in all domains has the corresponding content information denoted by $D_i$.  Different from the document with a common structure of the linguistic sequence, $D_i$ is a set of tags. \textbf{Note that all tags come from the same vocabulary\footnote{Dianping constructs a large-scale tag database extracted from user-generated contents via a customized tokenizer and provides a unified API to associate tags with contents.}, which is shared within all domains}.

The goal of TCDR is to improve the recommendation performance in the target domain by exploring latent tag-based interests behind cross-domain behaviors, which is especially needed for $U^{cs}$ and $U^{ia}$. In this work, we target at the top-K recommendation task\footnote{Dianping review feed recommender system is a complex industrial system with a tandem structure. After the top-K rating, also known as the matching or recall process, a customized CTR module will rate all candidate items by taking into account a cluster of additional features, like user age, occupation, and whether paid for advertising. Therefore, this work just aims to improve the recall performance.}.

\subsection{Graph Construction}
\begin{figure}[t]
	\centering
	\includegraphics[width=\linewidth]{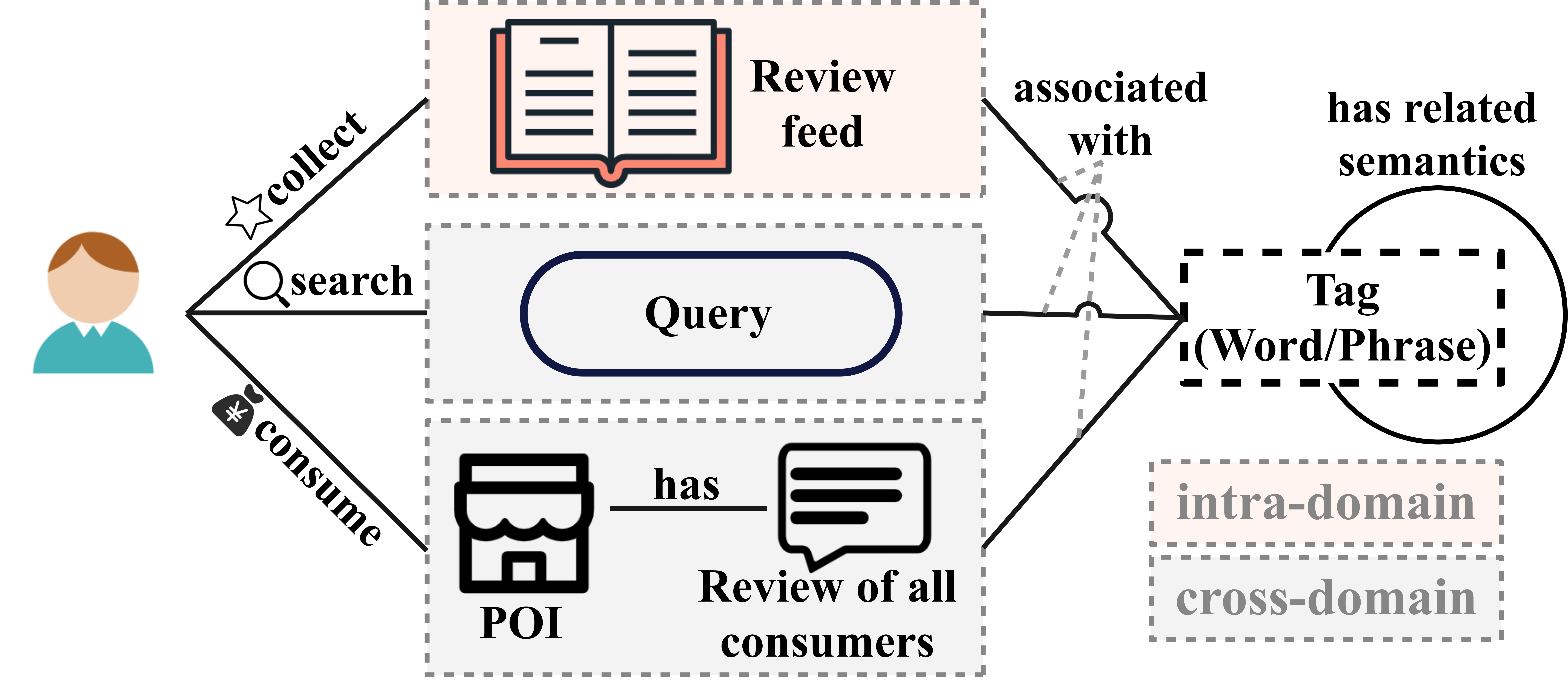}
	\caption{The tag-associated heterogenous graph constructed by Dianping data. Best viewed in color.}
	\Description{The heterogenous graph constructed by Dianping data.}
	\label{fig:graph_raw}
\end{figure}

\textbf{Definition 2.1. Tag-associated Heterogenous Graph.} 
Figure \ref{fig:graph_raw} illustrates three kinds of interactions from the review feed, search, and consumption domain in Dianping APP. There are six kinds of entities in this TCDR task, which are user, review feed, search query, POI, review of POI, and tag. We denote them as $\mathcal{U}=\{u\}$,$\mathcal{R}=\{r\}$,$\mathcal{Q}=\{q\}$,$\mathcal{P}=\{p\}$, $\mathcal{R}_\mathcal{P}=\{r_p\}$ and $\mathcal{T}=\{t\}$, respectively. Obviously, $i^t$ is identical to $r$, $i^s=\{i^{s_1},i^{s_2}\}=\{q,p\}$ and $D_i=\{t\}$. Eight relations are used to connect the above nodes, which are $\{u-r, u-q, u-p, p-r_p, r-t, q-t, r_p-t, t-t\}$. The first seven edges are easy to understand and we explain the last relation. $t-t$ represents two tags that have related semantics, like \textit{Japanese barbecue} and \textit{Japanese coffee}. We use Elasticsearch (ES) to store all tags and utilize its search engine to construct $t-t$ by regarding each tag as a query. Then a heterogeneous graph $\mathcal{G}$ is defined as $\mathcal{G}=\{ \mathcal{V}, \mathcal{E}\}$ with nodes $\mathcal{V}$ and edges $\mathcal{E}$. In Dianping dataset, $\mathcal{V}=\{\mathcal{U},\mathcal{R},\mathcal{Q},\mathcal{P},\mathcal{R}_\mathcal{P},\mathcal{T}\}$ and $\mathcal{E}=\{u-r, u-q, u-p, p-r_p, r-t, q-t, r_p-t, t-t\}$. Note that only $t-t$ is a symmetric relation and the others are directed. 

\textbf{Definition 2.2. Domain-specific Metapath.}
A metapath $\rho$ is a path defined on the graph $\mathcal{G}$, and is denoted in the form of ordered nodes$A_1 \rightarrow A_2 \rightarrow ...  \rightarrow A_{L}$. $d_{\rho}$ is the number of metapaths. In Dianping dataset, three are three metapaths ($d_{\rho}=3$) to represent domain-specific knowledge. $u-r-t$ represents user interested tags from the collected review feed. $u-q-t$ contains user potential preference inferred from his search log. $u-p-r_p-t$ describes what content of a local business can develop users' willingness to consume. 

\textbf{Definition 2.3. Metapath-based neighbor.}
Given a metapath $\rho$, the metapath-based neighbors $\mathcal{N}^{\rho}_{v}$ of a node $v$ is defined as the set of nodes that connect with $v$ using $\rho$. 

\section{The M2GNN framework}
\begin{figure*}[t]
	\centering
	\includegraphics[width=\linewidth]{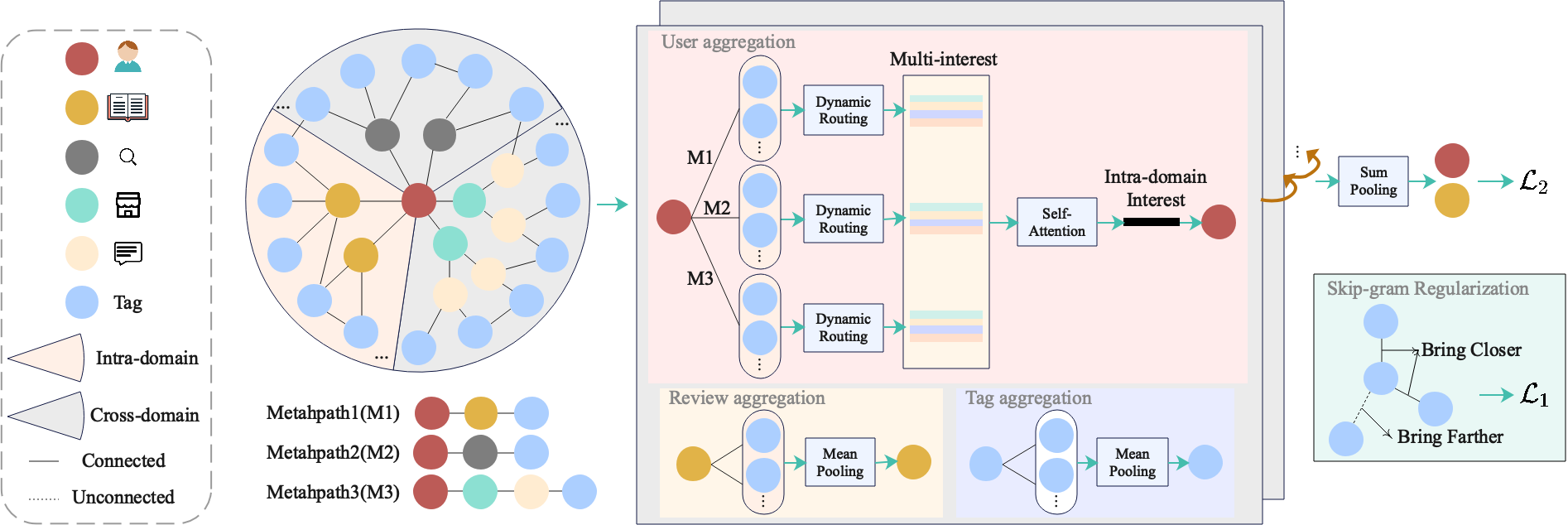}
	\caption{The framework of M2GNN. Best viewed in color.}
	\Description{The framework of M2GNN.}
	\label{fig:M2GNN}
\end{figure*}

\subsection{Motivation} \label{sec:motivaiton}

As aforementioned in \textbf{Q1}, the content side information in this TCDR task is tags. Different from contents with the linguistic sequence, tags have no natural sequential distribution, leading to the recent works focusing on sequential feature extraction like CATN, which will be compared in Section \ref{sec:expirements}, is less applicable. What's more, tags have an important feature, that is global shareability, which means any items in all domains are described by the same tag dict. This inspires us to recall the review feed including the same or similar tags appeared in other domains. For example, $user \xrightarrow{search} korean \, barbecue \xrightarrow{associated  \, with} barbecue   \xleftrightarrow{has  \, related  \, semantics} janpanese  \, barbecue \xleftarrow {associated  \, with} review \, feed$ can help to provide a review about janpanese barbecue for a user who searches korean barbecue. GNN, an intuitive tool to explore such a high-order reasoning chain, motivates us to build a tag-based graph and corresponding neural networks.

With determining to employ the GNN framework, we face severe noisy information in aggregation as shown in \textbf{Q2}. This is an inevitable problem when the number of attention members is large, which also appears in GraphAir \cite{hu2021graphair}. Hard attention \cite{xu2015show}, also known as nondeterministic attention, is a close solution to focus on exactly one or a few members, which will be compared in Section \ref{sec:expirements}. However, the tag is a  fine-grained object, which means a cluster of related tags indicates one entire high-level user preference. Therefore, discarding most of tags directly adopted in hard attention will lose content side information and degrade performance. Inspired by the success of multi-interest recommender systems (RSs), we find \textbf{extracting multiple interests from each domain can distill intra-domain interests in a corpus where noisy contents account for the most} with two reasons: (1) learning different aspects of user preference can preserve main but filter out trivial tags; (2) attention mechanism works better in transferring high-level interests into target domain than transferring tags directly because the number of interests is much less than that of tags.

\subsection{Overview}
A general framework that can generalize most GNN-based RSs (e.g. NGCF \cite{wang2019neural}, LightGCN \cite{he2020lightgcn}, KGAT \cite{wang2019kgat}, KGIN \cite{wang2021learning}, GC-SAN \cite{xu2019graph} and SURGE \cite{chang2021sequential}) adopts the following scheme: aggregate feature (neighbor) nodes to user or item (center) nodes and update $\rightarrow$ pool node representations after multiple layers $\rightarrow$ train the model according to downstream tasks like top-K recommendation or CTR prediction. 

Inspired by the above straightforward but effective idea, M2GNN aggregates tag nodes into user nodes by domain-specific metapaths without self-connection, learns item embeddings by their corresponding tags, and enhances representations of tag nodes by their homogeneous neighbors with related semantics. Since the number of tags associated with an item is small, adopting complex aggregation methods is unnecessary and here we use average pooling. The inherence of the tag-to-tag update is to bring the vectors of two tags with a similar meaning closer in latent semantic space, we also utilize average pooling and leave other aggregation functions like attention as a future work. Then sum operation is conducted to pool multi-layer embeddings, which is widely used in GNN for the recommendation task to fuse multi-hop collaborative filtering signals. Finally, the inner product is adopted to predict top-K items. Formally, the above message passing and recommendation mechanism are defined as follows,
\begin{equation}
	\label{eq:M2GNN-aggregate0}
	\begin{aligned}
		\mathbf{e}_{t}^{l+1}  \leftarrow \mathop{ \bm{{\rm Mean}} }\limits_{t^{\prime} \in \mathcal{N}^{t-t}_{t}} (\mathbf{e}_{{t^{\prime}}}^{l} )
	\end{aligned}
\end{equation}
\begin{equation}
	\label{eq:M2GNN-aggregate1}
	\begin{aligned}
		\mathbf{e}_{r}^{l+1}  \leftarrow \mathop{ \bm{{\rm Mean}} }\limits_{t \in \mathcal{N}^{r-t}_{r}} (\mathbf{e}_{{t}}^{l} )
	\end{aligned}
\end{equation}
\begin{equation}
	\label{eq:M2GNN-aggregate2}
	\begin{aligned}
		\mathbf{e}_{u}^{l+1}  \leftarrow \mathop{ \bm{{\rm Aggregate}} }\limits_{t \in \mathcal{N}^{\rho}_{u}} (\mathbf{e}_{t}^{l} )
	\end{aligned}
\end{equation}
\begin{equation}
	\label{eq:M2GNN-aggregate3}
	\begin{aligned}
		\mathbf{e}^{*}_v=\sum_{l=0}^{L}\mathbf{e}^{l}_{v} 
	\end{aligned}
\end{equation}
\begin{equation}
	\label{eq:M2GNN-aggregate4}
	\begin{aligned}
		y(u,r)={\mathbf{e}^{*}_u}^\top\mathbf{e}^{*}_r 
	\end{aligned}
\end{equation}
where $v$ can be either a user or an item node in target domain $r$, $\mathbf{e}_{v}^{l} \in  \mathbb{R}^{d \times 1}$ denotes the representation of node $v$ in the $l$-th layer, superscript $*$ indicates the final representation for predicting, $y(u,r)$ is the matching score of a pair of user and item. 

Obviously, the key component is equation \ref{eq:M2GNN-aggregate2}. As mentioned in Section \ref{sec:motivaiton}, we first learn multiple interests in each domain (cf. Section \ref{sec:IntraDA}) and transfer them into the target domain (cf. Section \ref{sec:InterDA}).

\subsection{Intra-domain Aggregation} \label{sec:IntraDA}
We utilize the dynamic routing network from \cite{sabour2017dynamic} as multi-interest extractor. For each domain (metapath), tag embeddings and user preferences are regarded as low and high-level capsules, respectively. Given a set of metapath-based neighbor tag nodes $t \in \mathcal{N}^{\rho}_{u}$, we learn interest capsules as
\begin{equation}
	\label{eq:drn}
	\begin{aligned}
			\mathbf{c}_{u,j}^{\rho,l}  = \sum_{t \in \mathcal{N}^{\rho}_{u}} w_{tj} \mathbf{S}  \mathbf{e}_{t}^{l}
		\end{aligned}
\end{equation}
where $\mathbf{c}_{u,j}^{\rho,l} $ denotes $j$-th interest capsule of user $u$ using metapath $\rho$ in $l$-th layer, $w_{tj}$ represents the contribution degree of low-level capsule $t$ to high-level capsule $j$, $\mathbf{S} \in  \mathbb{R}^{d \times d} $ is transformation matrix and is shared across all interest capsules and metapaths.

The superiority of dynamic routing is to update $w_{tj}$ via a new parameter called routing logit $b_{tj}$ in an iterative way. We initialize $b_{tj}$ randomly in the first iteration. $w_{tj}$  is calculated by performing softmax on $b_{tj}$ as
\begin{equation}
	\label{eq:drn1}
	\begin{aligned}
		w_{tj}=\frac{{\rm exp} \ b_{tj} }{\sum_{z=1}^{K} {\rm exp} \  b_{tz}}
	\end{aligned}
\end{equation}
where $K$ is the number of interest capsules and is adaptively calculated by the number of interacted tags as
\begin{equation}
	\label{eq:drnK}
	\begin{aligned}
			K={\rm max}(1, {\rm min}(K_{\rm max}, \lceil \log_2(| \mathcal{N}^{\rho}_{u} |) \rceil ))
		\end{aligned}
\end{equation}
where $K_{\rm max}$ is the maximum number of interest capsules and $\lceil x \rceil$ refers to the smallest integer that is more than or equal to $x$. 

To normalize the length of the output vector of a capsule between 0 and 1, which benefits updating $b_{tj}$ in next iteration, a non-linear squashing function is utilized as
\begin{equation}
	\label{eq:drn3}
	\begin{aligned}
		\mathbf{z}_{u,j}^{\rho,l}  = {\rm squash}(\mathbf{c}_{u,j}^{\rho,l} )=\frac{\left\|\mathbf{c}_{u,j}^{\rho,l}\right\|^{2}}{1+\left\|\mathbf{c}_{u,j}^{\rho,l}\right\|^{2}} \frac{\mathbf{c}_{u,j}^{\rho,l}}{\left\|\mathbf{c}_{u,j}^{\rho,l}\right\|}
	\end{aligned}
\end{equation}

Then $b_{tj}$ is updated by
\begin{equation}
	\label{eq:drn4}
	\begin{aligned}
		b_{tj} =b_{tj}+   \mathbf{z}_{u,j}^{\rho,l} \mathbf{S} \mathbf{e}_{t}^{l} 
	\end{aligned}
\end{equation}

Finally, the above routing process is repeated multiple times, which is set to be three, to obtain the converged output of interest capsules.

\subsection{Inter-domain Aggregation}\label{sec:InterDA}
We collect all interest embeddings as a matrix $\mathbf{Z}_{u}^{l}=[\mathbf{z}_{u,1}^{\rho_1,l},...,\mathbf{z}_{u,K_{\rm max}}^{\rho_1,l},\mathbf{z}_{u,1}^{\rho_2,l},...\mathbf{z}_{u,K_{\rm max}}^{\rho_{d_{\rho}},l}] \in \mathbb{R}^{d \times (d_{\rho} \cdot K_{\rm max})} $. For faster and unified tensor calculation, we pad zero vectors when $K < K_{\rm max}$. Now each user has an interest matrix ($\mathbf{Z}_{u}^{l}$) consisting of many interest embeddings ($\mathbf{z}_{u,j}^{\rho,l}$). To transfer useful preferences to the target domain, we assign a weight $\alpha_{u,j}^{\rho,l}$ to each interest embedding and output an intra-domain interest embedding by pooling $\mathbf{Z}_{u}^{l}$ attentively.  

Since sparse data in the target domain is undesirable to train accurate weights, a reasonable alternative solution is to utilize user similarity. For example, if two users have the same search and consumption log, it is reasonable to recommend a review feed collected by one user to the other. Technologically, we aim to assign the same weight to a pair of close interest embeddings from two users with similar portraits (a.k.a. $\mathbf{Z}_{u}^{l}$). Therefore, a global shared mechanism is required. To achieve this, a self-attention module is designed as
\begin{equation}
	\label{eq:SA1}
	\begin{aligned}
		\mathbf{H}_{u}^{l}=\mathbf{S}_{2}^{\top} \tanh \left(\mathbf{S}_{1} \mathbf{Z}_{u}^{l}\right) 
	\end{aligned}
\end{equation}
where $\mathbf{H}_{u}^{l}  \in \mathbb{R}^{1 \times (d_{\rho} \cdot K_{\rm max})} $ is the attention vector, $\mathbf{S}_1 \in \mathbb{R}^{d \times d}$ and $\mathbf{S}_2 \in \mathbb{R}^{d \times 1}$ are trainable matrixes, which are shared across all users. 

\textbf{Q2} reveals that only a fraction of interests contributes to the target domain, therefore we introduce the element-wise exponentiation operator before a softmax function to calculate weights as:
\begin{equation}
	\label{eq:SA1}
	\begin{aligned}
		\alpha_{u,j}^{\rho,l} =\frac{{\rm exp}({\rm pow} (h_{u,j}^{l},\gamma) )}{\sum_{z=1}^{d_{\rho} \cdot K_{\rm max}} {\rm exp} ({\rm pow}( h_{u,z}^{l},\gamma)) }
	\end{aligned}
\end{equation}
where $h_{u,j}^{l}$ is the $j-$th element in $\mathbf{H}_{u}^{l}$ and $\gamma$ is a hyperparameter to control the attention distribution. As $\gamma$ increases, more significant interest embeddings will receive more attention and bigger weights, and vice versa. Consider the limit case, $\gamma=0$ is identical to mean pooling and $\gamma= \infty $ means the hard attention which chooses only one candidate as output. Therefore, $\gamma$ is set to be bigger than 1, which can alleviate the problem in \textbf{Q2}.

Finally, the user representation is inferred to represent his intra-domain interests as
\begin{equation}
	\label{eq:SA2}
	\begin{aligned}
		\mathbf{e}_{u}^{l+1}=  \mathbf{Z}_{u}^{l} {\mathbf{A}_{u}^{l}}^{\top} 
	\end{aligned}
\end{equation}
where $\mathbf{A}_{u}^{l}=[\alpha_{u,1}^{\rho,l},...,\alpha_{u,d_{\rho} \cdot K_{\rm max}}^{\rho,l}] \in  \mathbb{R}^{1 \times (d_{\rho} \cdot K_{\rm max})}$ is the weight matrix of user $u$ in $l$ layer.

\subsection{Skip-gram Regularization}\label{sec:SGR}
It is clear that the quality of tag embeddings deeply affects the clustering performance in Section \ref{sec:IntraDA} and weight computing in Section \ref{sec:InterDA}. To further draw related tags closer and unconcerned tags removed in latent semantic space, we adopt skip-gram idea to regularize the training process. Specifically, a tag node is regarded as the center word, its homogeneous neighbor nodes are set to be nearby words. The principle is to maximize the co-occurrence probability of related tags, and vice versa. We randomly sample a negative node for each positive pair of tags and define skip-gram objective loss as 
\begin{equation}
	\label{eq:SGr}
	\begin{aligned}
		\mathcal{L}_1 =-\frac{1}{|\mathcal{E}^{t-t}|} \sum_{t \in \mathcal{T}} \sum_{t^{\prime} \in \mathcal{N}^{t-t}_{t}, t^{\prime \prime} \notin \mathcal{N}^{t-t}_{t}} ( p(t^{\prime} \mid t)-p(t^{\prime\prime} \mid t) )
	\end{aligned}
\end{equation}
where $|\mathcal{E}^{t-t}|$ denotes the number of all edges belonging to type $t-t$ in graph $\mathcal{G}$.

The basic formulation to model probability $p$ is the global softmax performed on all tags, which is impractical for the huge cost of time and calculation. Inspired by hierarchical softmax, an auxiliary set of trainable embeddings $\mathcal{V} \in \mathbb{R}^{d \times |\mathcal{T}|}$ is introduced to define $p$ as
\begin{equation}
	\label{eq:SGr}
	\begin{aligned}
		p(t^{\prime} \mid t)=\operatorname{log} \sigma({\mathbf{v}_{t^{\prime}}}^{\top} \mathbf{e}_{t}^{0})
	\end{aligned}
\end{equation}
where $\mathbf{v}_{t^{\prime}}$ is the embedding of tag $t^{\prime}$ from $\mathcal{V}$, $ \mathbf{e}_{t}^{0} $ is the initial embedding of tag $t$.

\subsection{Model Optimization}
With the user representation $\mathbf{e}^{*}_u$ and the review representation $\mathbf{e}^{*}_r$ ready, equation \ref{eq:M2GNN-aggregate4} is adopted to predict top-K items. 
We employ the BPR loss \cite{rendle2012bpr} as
\begin{equation}
	\label{eq:loss2}
	\begin{aligned}
		\mathcal{L}_2=-\frac{1}{|\mathcal{O}|} \sum_{(u,r^{+},r^{-})\in \mathcal{O}} \ln \sigma \bigg( y(u,r^{+})-y(u,r^{-}) \bigg) +\lambda\|\Theta\|_{2}^{2}
	\end{aligned}
\end{equation}
where $\mathcal{O}=\{ (u,r^{+},r^{-}) | (u,r^{+}) \in \hat{y}_{ui^{t}}, (u,r^{-}) \notin \hat{y}_{ui^{t}}\}$ denotes the training set, where $r^{+}$ is the observed interaction and $r^{-}$ is a negative counterpart which is sampled from the unobserved interactions randomly, $\sigma(\cdot)$ is the sigmoid function. $\Theta=\{ \mathbf{e}_x^{0}, \mathcal{S},\mathcal{S}_1,\mathcal{S}_2, \mathbf{v}_t |x \in \{\mathcal{U},\mathcal{R},\mathcal{T}\},t \in \mathcal{T}\}$ is the model
parameter set. $L_2$ regularization parameterized by $\lambda$ on $\Theta$ is conducted to prevent overfitting. Finally, the joint loss is 
\begin{equation}
	\label{eq:loss}
	\begin{aligned}
		\mathcal{L}=\mathcal{L}_1+\mathcal{L}_2
	\end{aligned}
\end{equation}



%
%

\section{experiments} \label{sec:expirements}
We evaluate our proposed M2GNN method on two benchmark datasets to answer the following research questions: 
\begin{itemize}
	\item \textbf{RQ1}: Does our proposed M2GNN outperform the state-of-the-art recommendation methods?
	\item \textbf{RQ2}: How do different components (i.e., two-step aggregation module and skip-gram regularization) and hyperparameters (i.e. capsule numbers, GNN layer numbers, and exponent before softmax) affect M2GNN?
	\item \textbf{RQ3}: Does M2GNN achieve great improvement towards cold-start and inactive users?
	\item \textbf{RQ4}: Is there an online improvement of review feed recommendation in Dianping homepage by M2GNN?
	\item \textbf{RQ5}: Can M2GNN provide potential explanations about tag-based preferences of users?
\end{itemize}

\subsection{Experimental Settings}

\noindent\textbf{Dataset Description}. 
We propose a novel industrial dataset, called DPBJ (Dianping Beijing). The dataset is encrypted and desensitized, and does not contain any personally identifiable information. To avoid data leakage, we collect the multi-site data in one month (from Nov. 1, 2022, to Nov. 30, 2022) as the training and validation set. Then the latter one-week data (from Dec. 1, 2022, to Dec. 7, 2022) is the testing set. We also use a common dataset, Amazon review dataset\footnote{http://jmcauley.ucsd.edu/data/amazon/}, where the book and movie are chosen as the source and target domain. To ensure data quality, we remove the interaction records without review text and some low-quality items following earlier studies \cite{he2017neural,kang2019semi}. We split the review text into individual words and filter out stopwords by word tokenization tool like nltk\footnote{https://github.com/nltk/nltk}. The above all words comprise the tag dict. The detailed statistics of these two datasets are reported in Table \ref{tab:Statistics of two datasets}. 
\begin{table}[t]  
	\centering  
	\caption{Statistics of two datasets. $m$ and $b$ denote movie and book, respectively. 'int.' indicates the interaction data of each domain.}    
	\begin{tabular}{cccccc}   
		 \hline 
		 &   \multicolumn{1}{c|}{}    & \multicolumn{2}{c|}{DPBJ} & \multicolumn{2}{c}{Amazon} \\
		 \cline{1-6}
		  \multicolumn{2}{c|}{\# $u$} & \multicolumn{2}{c|}{6,156,837} & \multicolumn{2}{c}{41,465} \\
		  \cline{1-6}
		 \multicolumn{2}{c|}{\# $t$} & \multicolumn{2}{c|}{94,029} & \multicolumn{2}{c}{20,000} \\
		 \cline{1-6}
		 \multicolumn{2}{c|}{\# $t$-$t$} & \multicolumn{2}{c|}{1,679,832} & \multicolumn{2}{c}{184,776} \\
		 \cline{1-6}
		 \multirow{3}{*}{$D^{t}$} & \multicolumn{1}{|c|}{item}  & \# $r$   & \multicolumn{1}{c|}{2,036,665} & \# $m$   & 7,379 \\
		\cline{2-6}
		 & \multicolumn{1}{|c|}{\multirow{2}{*}{int.}} & \# $u$-$r$   & \multicolumn{1}{c|}{297,658} & \# $u$-$m$   & 109,864 \\
		 \cline{3-6}
		 & \multicolumn{1}{|c|}{}  & \# $r$-$t$   & \multicolumn{1}{c|}{2,917,242} & \# $m$-$t$   & 49,127,984 \\
		 \cline{1-6}
         \multirow{3}{*}{$D^{s_1}$} & \multicolumn{1}{|c|}{item}  & \# $q$   & \multicolumn{1}{c|}{3,436,125} & \# $b$   & 34,396 \\
    	 \cline{2-6}
		 & \multicolumn{1}{|c|}{\multirow{2}{*}{int.}} & \# $u$-$q$   & \multicolumn{1}{c|}{54,527,624} & \# $u$-$b$   & 1,425,569 \\
		 \cline{3-6}
		 & \multicolumn{1}{|c|}{}  & \# $q$-$t$   & \multicolumn{1}{c|}{5,936,250} & \# $b$-$t$   & 169,399,312 \\
		 \cline{1-6}
		 \multirow{5}{*}{$D^{s_2}$} & \multicolumn{1}{|c|}{\multirow{2}{*}{item}} & \# $p$   & \multicolumn{1}{c|}{207,766}    & \multicolumn{2}{c}{\multirow{5}{*}{}} \\
		 \cline{3-4}
		 &   \multicolumn{1}{|c|}{}     & \# $r_p$   & \multicolumn{1}{c|}{2,103,670}  & \multicolumn{2}{c}{} \\
		 \cline{2-4}
		 & \multicolumn{1}{|c|}{\multirow{3}{*}{int.}} & \# $u$-$p$  & \multicolumn{1}{c|}{15,473,732} & \multicolumn{2}{c}{} \\
		 \cline{3-4}
		 &  \multicolumn{1}{|c|}{}  & \# $p$-$r_p$   & \multicolumn{1}{c|}{160,263}  & \multicolumn{2}{c}{} \\
		 \cline{3-4}
		 &  \multicolumn{1}{|c|}{}  & \# $r_p$-$t$   & \multicolumn{1}{c|}{3,646,553}  & \multicolumn{2}{c}{} \\    
		 \hline
		 \multicolumn{2}{c|}{\multirow{3}{*}{metapath}} & \# $u$-$r$-$t$ & \multicolumn{1}{c|}{497,532}  & \# $u$-$m$-$t$ & 64,498,631 \\
		 \cline{3-6}
		 \multicolumn{2}{c|}{} & \# $u$-$q$-$t$ & \multicolumn{1}{c|}{73,904,447}  & \# $u$-$b$-$t$ & 354,326,927 \\
		 \cline{3-6}
		 \multicolumn{2}{c|}{} & \# $u$-$p$-$r_p$-$t$ & \multicolumn{1}{c|}{184,203,329}  & \multicolumn{2}{c}{} \\ 
		 \hline
	\end{tabular}  
\label{tab:Statistics of two datasets}
\end{table}

\begin{table*}[t]  
	\centering  \caption{Performance comparison. $\blacktriangle$\% denotes the relative improvement of M2GNN over the best SOTA algorithm.}    
	\begin{tabular}{cccccccccc}          
		\hline  
		\multicolumn{2}{c|}{\multirow{2}{*}{}} & \multicolumn{4}{c|}{DPBJ}     & \multicolumn{4}{c}{Amazon} \\
		\cline{3-10}
		\multicolumn{2}{c|}{} & 
		\multicolumn{1}{c}{Recall@50} & \multicolumn{1}{c}{Recall@100} & \multicolumn{1}{c}{Hit@50} & \multicolumn{1}{c|}{Hit@100} & 
		\multicolumn{1}{c}{Recall@50} & \multicolumn{1}{c}{Recall@100} & \multicolumn{1}{c}{Hit@50} & \multicolumn{1}{c}{Hit@100} \\
		\hline
		\multicolumn{1}{c|}{\multirow{2}{*}{single-domain}} & \multicolumn{1}{c|}{MF} & 
		\multicolumn{1}{c}{0.0265} & \multicolumn{1}{c}{0.0421} & \multicolumn{1}{c}{0.0529} & \multicolumn{1}{c|}{0.0796} & 
		\multicolumn{1}{c}{0.0612} & \multicolumn{1}{c}{0.0989} & \multicolumn{1}{c}{0.3874} & \multicolumn{1}{c}{0.5218} \\
		\multicolumn{1}{c|}{} & \multicolumn{1}{c|}{LightGCN} & 
		\multicolumn{1}{c}{0.0388} & \multicolumn{1}{c}{0.0595} & \multicolumn{1}{c}{0.0697} & \multicolumn{1}{c|}{0.1045} & 
		\multicolumn{1}{c}{\underline{0.1044}} & \multicolumn{1}{c}{0.1599} & \multicolumn{1}{c}{0.5232} & \multicolumn{1}{c}{0.6475} \\
		\cline{1-10}
		\multicolumn{1}{c|}{\multirow{2}{*}{\minitab[c]{content-free \\ cross-domain}}} & \multicolumn{1}{c|}{EMCDR} & 
		\multicolumn{1}{c}{0.0425} & \multicolumn{1}{c}{0.0699} & \multicolumn{1}{c}{0.0752} & \multicolumn{1}{c|}{0.1137} & 
		\multicolumn{1}{c}{0.0670} & \multicolumn{1}{c}{0.1056} & \multicolumn{1}{c}{0.4271} & \multicolumn{1}{c}{0.5790} \\
		\multicolumn{1}{c|}{} & \multicolumn{1}{c|}{PTUPCDR} & 
		\multicolumn{1}{c}{0.0499} & \multicolumn{1}{c}{0.0731} & \multicolumn{1}{c}{0.0825} & \multicolumn{1}{c|}{0.1331} & 
		\multicolumn{1}{c}{0.0727} & \multicolumn{1}{c}{0.1341} & \multicolumn{1}{c}{0.4901} & \multicolumn{1}{c}{0.6515} \\
		\cline{1-10}
		\multicolumn{1}{c|}{\multirow{2}{*}{\minitab[c]{content-based \\ cross-domain}}} & \multicolumn{1}{c|}{CATN} & 
		\multicolumn{1}{c}{0.0304} & \multicolumn{1}{c}{0.0495} & \multicolumn{1}{c}{0.0560} & \multicolumn{1}{c|}{0.0829} & 
		\multicolumn{1}{c}{0.0815} & \multicolumn{1}{c}{0.1354} & \multicolumn{1}{c}{0.5323} & \multicolumn{1}{c}{0.6811} \\
		\multicolumn{1}{c|}{} & \multicolumn{1}{c|}{MTNet} & 
		\multicolumn{1}{c}{\underline{0.0531}} & \multicolumn{1}{c}{\underline{0.0822}} & \multicolumn{1}{c}{0.0877} & \multicolumn{1}{c|}{0.1362} & 
		\multicolumn{1}{c}{0.0707} & \multicolumn{1}{c}{0.1199} & \multicolumn{1}{c}{0.4761} & \multicolumn{1}{c}{0.6377} \\
		\cline{1-10}
		\multicolumn{1}{c|}{\multirow{3}{*}{\minitab[c]{GNN-based \\ cross-domain} } } & \multicolumn{1}{c|}{HeroGRAPH} & 
		\multicolumn{1}{c}{0.0437} & \multicolumn{1}{c}{0.0713} & \multicolumn{1}{c}{\underline{0.0884}} & \multicolumn{1}{c|}{\underline{0.1381}} & 
		\multicolumn{1}{c}{0.0884} & \multicolumn{1}{c}{0.1523} & \multicolumn{1}{c}{0.5514} & \multicolumn{1}{c}{0.7065} \\
		\multicolumn{1}{c|}{} & \multicolumn{1}{c|}{HCDIR} & \multicolumn{1}{c}{0.0320} & \multicolumn{1}{c}{0.0562} & \multicolumn{1}{c}{0.0653} & \multicolumn{1}{c|}{0.1099} & 
		\multicolumn{1}{c}{0.1005} & \multicolumn{1}{c}{0.1680} & \multicolumn{1}{c}{0.5785} & \multicolumn{1}{c}{0.7194} \\
		\multicolumn{1}{c|}{} & \multicolumn{1}{c|}{GA-MTCDR-P} & \multicolumn{1}{c}{0.0341} & \multicolumn{1}{c}{0.0577} & \multicolumn{1}{c}{0.0690} & \multicolumn{1}{c|}{0.1153} & 
		\multicolumn{1}{c}{0.1028} & \multicolumn{1}{c}{\underline{0.1703}} & \multicolumn{1}{c}{\underline{0.5977}} & \multicolumn{1}{c}{\underline{0.7318}} \\
		\cline{1-10}
		\multicolumn{2}{c|}{M2GNN} &
		\multicolumn{1}{c}{\textbf{0.0629}} & \multicolumn{1}{c}{\textbf{0.0953}} & \multicolumn{1}{c}{\textbf{0.0967}} & \multicolumn{1}{c|}{\textbf{0.1464}} & 
		\multicolumn{1}{c}{\textbf{0.1115}} & \multicolumn{1}{c}{\textbf{0.1783}} & \multicolumn{1}{c}{\textbf{0.6471}} & \multicolumn{1}{c}{\textbf{0.7768}} \\
		\hline    
		\multicolumn{2}{c|}{$\blacktriangle$\%} &
		\multicolumn{1}{c}{18.46} & \multicolumn{1}{c}{{15.94}} & \multicolumn{1}{c}{{9.39}} & \multicolumn{1}{c|}{{6.01}} & 
		\multicolumn{1}{c}{{6.80}} & \multicolumn{1}{c}{{4.70}} & \multicolumn{1}{c}{{8.27}} & \multicolumn{1}{c}{{6.15}} \\
		\hline  
	\end{tabular}
	\label{tab:Overall Performance Comparison}
\end{table*}

~\\
\noindent\textbf{Baselines}. 
To demonstrate the effectiveness, we compare M2GNN with \textbf{single-domain} (MF, lightGCN), \textbf{content-free cross-domain} (EMCDR, PTUPCDR), \textbf{content-based cross-domain} (CATN, MTNet), and \textbf{GNN-based cross-domain} (HeroGRAPH, HCDIR, GA) methods. 
\begin{itemize}
	\item \textbf{MF} \cite{rendle2012bpr} is a matrix factorization method optimized by the Bayesian personalized ranking (BPR) loss. 
	\item \textbf{lightGCN} \cite{he2020lightgcn} simplifies NGCF \cite{wang2019neural} and provides simple but effective model for user-item bipartite graph-based RS.
	\item \textbf{EMCDR} \cite{man2017cross} first pretrains embeddings in each domain and then learns a mapping function to transfer interests across domains.
	\item \textbf{PTUPCDR} \cite{zhu2022personalized} designs a meta network fed with users’ characteristic embeddings to generate personalized bridge functions to achieve personalized transfer of user preferences.
	\item \textbf{CATN} \cite{wang2019kgat} extracts multiple aspects from the documents and learns aspect correlations across domains attentively.
	\item \textbf{MTNet} \cite{hu2018mtnet} designs a memory component to find the words highly relevant to the user preferences, which are then fused with source domain-specific knowledge to improve final prediction.
	\item \textbf{HeroGRAPH} \cite{cui2020herograph} constructs a heterogeneous graph to fuse all interactions as a whole and learn additional global embeddings to enhance recommendation in all domains.
	\item \textbf{HCDIR} \cite{bi2020heterogeneous} separately learns user embeddings in each domain via an insurance information network and then employs MLP to perform latent space matching across domains.
	\item \textbf{GA} \cite{zhu2021unified} proposes a unified framework, which constructs separate heterogeneous graphs to generate more representative embeddings and uses an element-wise attention mechanism to effectively combine them of common entities.
\end{itemize}

\noindent\textbf{Parameter Settings}. 
We implement our M2GNN model in Pytorch and Deep Graph Library (DGL)\footnote{https://github.com/dmlc/dgl}, which is a Python package for deep learning on graphs. Since DPBJ has million-scale nodes and billion-scale edges, the API of distributed data parallel (DDP) in Pytorch is used to train M2GNN and the data loading package in DGL is adopted to generate subgraph samples for each batch. We released all implementations (code, datasets, parameter settings, and training logs) to facilitate reproducibility\footnote{https://github.com/huaizepeng2020/M2GNN\_git}. The embedding size of all models is set to be 64 for a fair comparison. We adopt Adam \cite{kingma2014adam} as the optimizer and the batch size is fixed at 1024 for all methods. 

~\\
\noindent\textbf{Evaluation Metrics}. 
As mentioned in Section \ref{sec:problemsetup}, Recall@$K$ and Hit@$K$\footnote{K is set to be 50 rather than a commonly used value like 20 or 100. The reason is that the subsequent CTR module only ranks the top 50 items from the recall phase.} are used in the task of top-K recommendation. 

\subsection{Performance Comparison (RQ1)}
We report the empirical results in Table \ref{tab:Overall Performance Comparison}. The observations are as follows:
\begin{itemize}
	\item M2GNN consistently achieves the best performance on two datasets in terms of all measures. Specifically, it achieves significant improvements over the strongest baselines w.r.t. Recall@50 by 18.46\%, 6.80\% in DPBJ and Amazon, respectively. The results prove the effectiveness of M2GNN. 
	\item Despite some particular cases, the average performances of content-aware (content-based\&GNN-based) cross-domain, content-free cross-domain, and single-domain models are in decreasing order. A clear reason is whether to introduce the auxiliary information, like contents or cross-domain interactions. 
	\item MTNet outperforms CATN in DPBJ while underperforms CATN in Amazon. The reason is the difference between content extractors in these two models. MTNet adopts an attention network to model unstructured text without regarding the sequential feature, thus performing better in DPBJ dataset. CATN utilizes text convolution and has superiority in modeling sequential signals, which is more suitable for the Amazon dataset.
	\item HeroGRAPH outperforms GA-MTCDR-P (HCDIR) in DPBJ while underperforms GA-MTCDR-P (HCDIR) in Amazon. DPBJ has a large number of noisy tags, leading to a big gap between intra and cross-domain interests. Therefore, using multiple embeddings to represent domain-specific interests separately (HeroGRAPH) performs better than using one representation (GA-MTCDR-P/HCDIR). However, Amazon dataset has higher consistency in user preferences toward books and movies, thus aggregating cross-domain interests is more applicable.
\end{itemize}
\begin{table}[t]  
	\centering  
	\caption{Impact of two-step aggregation and skip-gram regularization.}   
	\begin{tabular}{c|cc|cc}
		\hline
		\multirow{2}{*}{} & \multicolumn{2}{c|}{DPBJ} & \multicolumn{2}{c}{Amazon} \\
		\cline{2-5}
		& Recall@50 & Hit@50 & Recall@50 & Hit@50 \\ 
		\hline
		M2GNN-mean & 0.0511     & 0.0762     & 0.0864     & 0.5264 \\
		M2GNN-softmax & 0.0524     & 0.0791     & 0.0837     & 0.4879 \\
		M2GNN-HAN & 0.0544     & 0.0809     & 0.0898     & 0.5631 \\
		M2GNN-hard & 0.0586     & 0.0873     & 0.0951     & 0.6062 \\ 
		\hline\hline
		M2GNN-w/o$\mathcal{L}_1$& 0.0526     & 0.0822     & 0.0903     & 0.5400 \\
		\hline
	\end{tabular}  
	\label{tab:Impact of two}
\end{table}

\subsection{Study of M2GNN (RQ2)}
\begin{figure*}[t]
	\centering
	\vspace{-0.35cm} 
	\subfigtopskip=0pt 
	\includegraphics[width=0.75\linewidth]{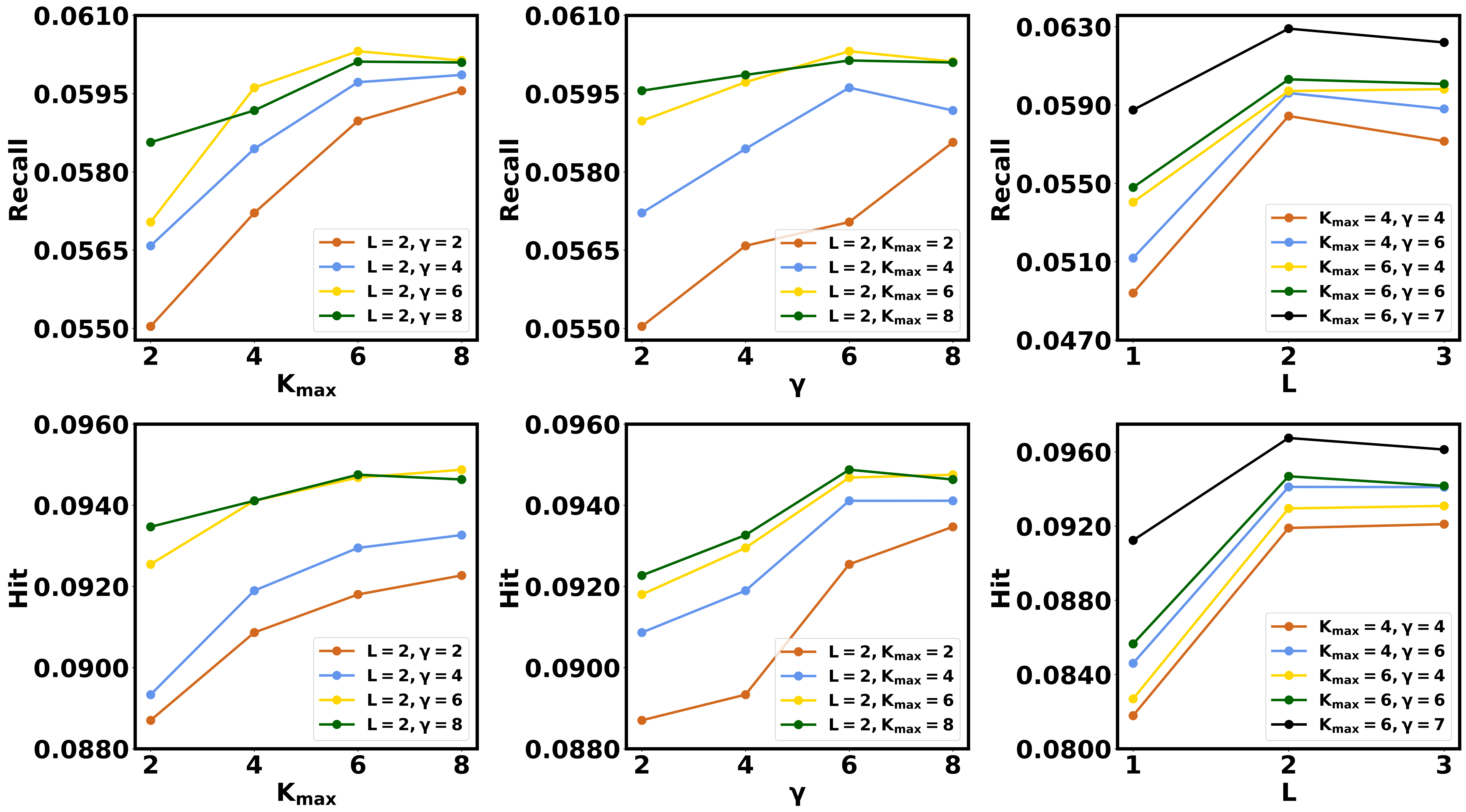}
	\caption{Effect of hyperparameters. Best viewed in color.}
	\Description{The impact of hyperparameters.}
	\label{fig:impact of hyperparameters}
\end{figure*}
\begin{figure}[t]
	\centering
	\vspace{-0.35cm} 
	\subfigtopskip=0pt 
	\includegraphics[width=0.95\linewidth]{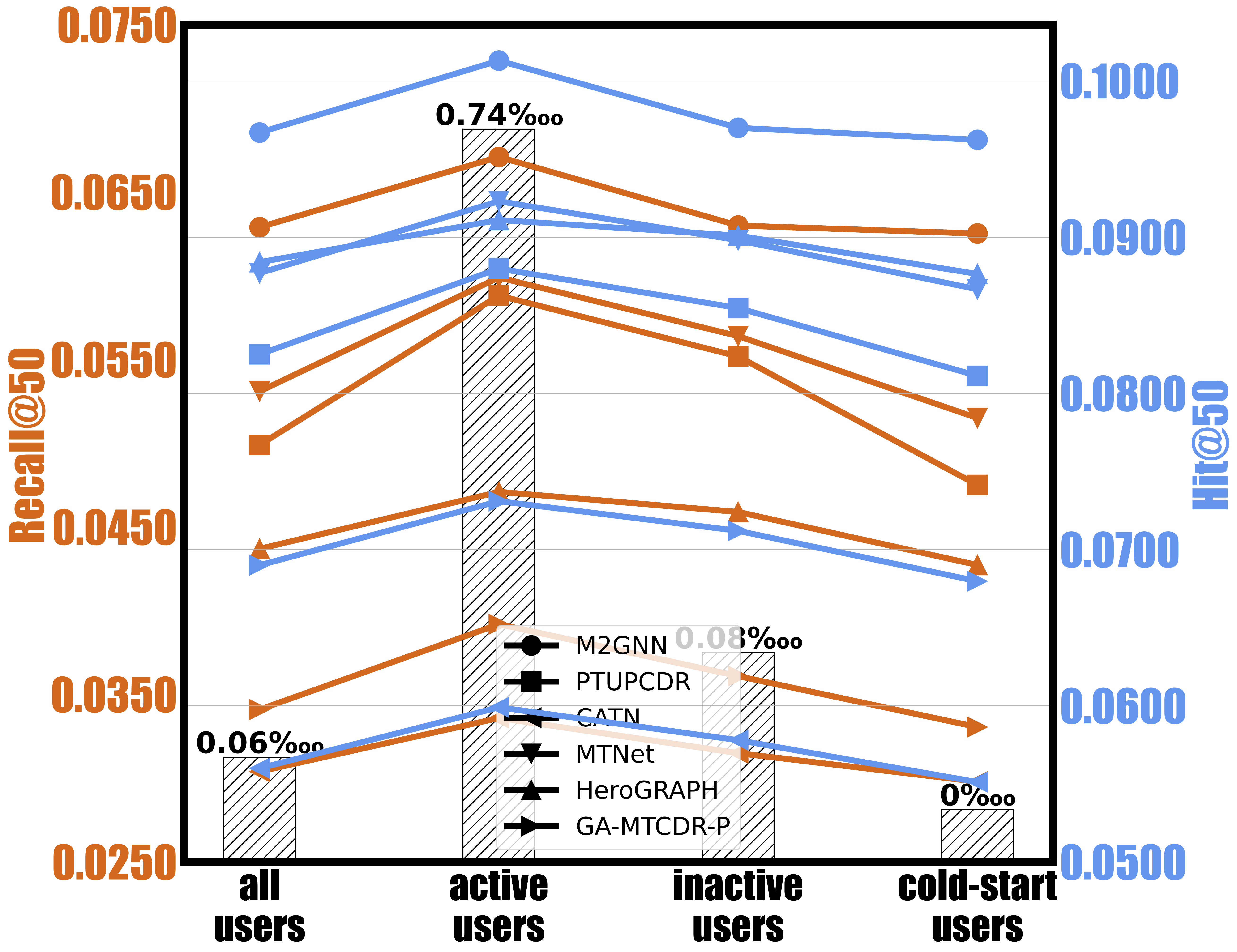}
	\caption{Performance comparison over the sparsity distribution of user groups. The background
		histograms indicate the density of each user group and ‱ means permyriad.}
	\Description{The impact of hyperparameters.}
	\label{fig:impact of user_group}
\end{figure}

\noindent\textbf{Impact of two-step aggregation module}. To demonstrate the superiority of the two-step aggregation module, we compare the performance of M2GNN with the following four variants which adopt different methods of aggregation.
\begin{itemize}
	\item \textbf{M2GNN-mean} uses average pooling in both two steps.
	\item \textbf{M2GNN-softmax} adopts the standard schema of attention, proposed in Transformer \cite{vaswani2017attention}, in both two steps. $K$ is user embedding, while $Q$ and $V$ are tag embeddings.
	\item \textbf{M2GNN-HAN} regards intra-domain and inter-domain aggregation as node-level and semantic-level attention in HAN \cite{wang2019heterogeneous}.
	\item \textbf{M2GNN-hard} utilizes local alignment attention in \cite{luong2015effective}, which extends original hard attention \cite{xu2015show} to select a subset of input members via implementing a softmax distribution. 
\end{itemize}
The results are shown in the upper part of Table \ref{tab:Impact of two} and the major findings are as follows:
\begin{itemize}
	\item M2GNN-mean and M2GNN-softmax achieve the worst performances on the two datasets. The reason is both of them don't track the noisy content problem. The former regards all tags as equally important, while the latter receives a large number of attention members, leading to the weights of significant tags are not trained as expected.
	\item M2GNN-HAN slightly outperforms the first two variants. A possible reason is that the different mechanism to calculate attention scores using the type-specific transformation matrix and the weight coefficient helps to increase the weights of useful tags.
	\item M2GNN-hard performs the best among the four variants. A clear reason is that only a subset of significant tags are aggregated into user representations, which avoids using all tags to infer final embeddings. However, as aforementioned in Section \ref{sec:motivaiton}, it is inevitable to discard some useful tags in hard attention, which causes that M2GNN-hard underperforms M2GNN.
\end{itemize}

\noindent\textbf{Impact of skip-gram regularization}. 
To demonstrate the superiority of the skip-gram regularization, we discard $\mathcal{L}_1$ and only leverage collaborative filtering signals to train the model, termed  M2GNN-w/o$\mathcal{L}_1$. The lower part of Table \ref{tab:Impact of two} shows that M2GNN-w/o$\mathcal{L}_1$ sharply underperforms M2GNN without regularizing tag embeddings via $\mathcal{L}_1$ , which proves the necessity of the skip-gram regularization.


\noindent\textbf{Impact of hyperparameters}. 
We investigate the influence of three hyperparameters: layer numbers, capsule numbers, and exponent before softmax, all of which greatly affect multi-interest extraction and transfer. $L$, $K_{{\rm max}}$ and $\gamma$ are searched in $\{1,2,3\}$, $\{2,4,6,8\}$ and $\{2,4,6,8\}$, respectively. Figure \ref{fig:impact of hyperparameters} shows the results in the DPBJ dataset and it has a similar conclusion in the Amazon dataset, which is omitted. From the observations above, we make several conclusions. Firstly, $L=2$ results in the best performance, which means 2-order connections integrate more similar tags into users and stacking more layers continually brings in noisy tags, which is harmful to the recommendation. Secondly, the increase of interest capsules indeed improves the recall list since local businesses contain multiple aspects, while there is a saturation limit which has a similar conclusion in \cite{li2019multi}. Thirdly, the optimal number of $\gamma$ is 6\textasciitilde8. The reason is that a proper exponent helps the softmax function to identify more important interests and assign bigger scores.
\begin{figure*}[t]
	\centering
	\includegraphics[width=\linewidth]{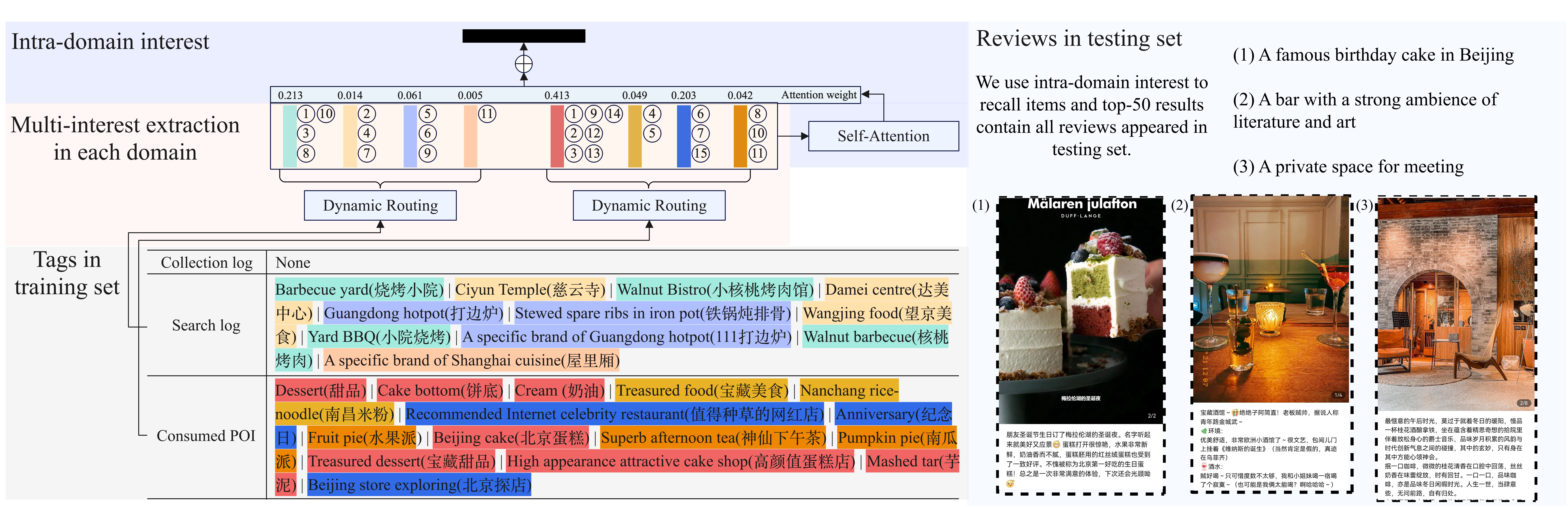}
	\caption{Case study. Each tag is represented by a color block. The original Chinese text is in brackets and we translate it into English. Best viewed in color.}
	\Description{Case study.}
	\label{fig:case_study}
\end{figure*}
\subsection{Cold-start and Inactive Users Analysis (RQ3)}
One motivation for exploiting CDR is to alleviate the severe sparsity problem in Dianping, where an overwhelming majority of users are cold-start in Figure \ref{fig:Data feature}. Therefore, we investigate whether introducing CDR helps alleviate this issue. Towards this end, we divide the testing set into four groups based on the interaction number per user, which has the same criterion in Section \ref{sec:introduction}.  Metrics are calculated for each group individually. EMCDR and HCDIR have similiar performances with PTUPCDR and GA-MTCDR-P respectively, therefore they are omitted. Figure \ref{fig:impact of user_group} illustrates the results and we have two conclusions: (1) All models perform better for active users since they have richer interactions in the target domain. (2) There is a marked decrease caused by the data sparsity problem for inactive and cold-start users in baselines, while M2GNN has the least impact. This proves M2GNN has stronger robustness to serve the inactive and cold-start users.

\subsection{Online A/B Test (RQ4)}
We report a 2-week online A/B test experiment of M2GNN in Beijing, from Dec. 1, 2022, to Dec. 15, 2022. Specifically,  the current recommender system on the Dianping homepage\footnote{This is not disclosed for confidentiality reasons.} is the base model. M2GNN is integrated into it as an additional recall method and directly offers 200 reviews for the later rank stage, which is regarded as the tested model. 3.4 million users participated in this A/B test. Three kinds of metrics are chosen. ‘Collection number’ (CN) indicates the number of all collected reviews, which is an absolute amount. ‘Collection number per user’ (UV-CN) means the number of collected reviews per user, which eliminates users with no collection behaviors. 'Collection rate' (CR) is the ratio of collected reviews to all exposed reviews. We just give the increase rate since the true value of each metric is sensitive. As in Table \ref{tab:Online}, M2GNN achieves great improvements relative to the base model. What's more, the improvement of inactive and cold-start users (1.1986\% CR) is more significant than that of active users (0.4663\% CR), which demonstrates the superiority of M2GNN in alleviating data sparsity problems by transferring cross-domain knowledge.
\begin{table}[t]
	\caption{Online metrics in a week.}
	\label{tab:Online}
	\begin{tabular}{c|ccc}
		\hline
		\multicolumn{1}{c|}{Metrics}  & CN& UV-CN & CR\\
		\hline\hline
		\multicolumn{1}{c|}{active users} & +0.4715\% & +0.3749\% & +0.4663\% \\
		inactive and  cold-start users & +1.0957\% & +1.3447\% & +1.1986\% \\
		\hline           
	\end{tabular}
\end{table}

\subsection{Case Study (RQ5)}
To prove the interpretability of M2GNN, we visualize the two-step aggregation process to show how to distill intra-domain interests from tags, most of which are noises. Figure \ref{fig:case_study} takes a specific user as an example. The bottom lists all tags from three domains, and obviously he is a cold-start user because of no interactions in the target domain. In the pink block, there are four interest capsules in each domain according to equation \ref{eq:drnK}. The number $\textcircled{x}$ near to a capsule $\mathbf{z}$ represents that $x$-th tag has the largest routing logit towards $\mathbf{z}$ among four capsules, which is also reflected by highlighting tags with the same color. The results indicate that tags are clustered via their semantics and the interest capsule represents the high-level preference of each domain. For example, in the consumption domain, the first capsule consists of tags about cakes (e.g. \textit{Dessert}, \textit{Cream} and \textit{Beijing cake}). What's more, the influence of trivial tags (e.g. \textit{Mashed tar}) is weakened since a capsule can only represent one aspect of semantics. Then eight weights are calculated by the self-attention module as shown in the blue block. Three capsules, which are concerned with barbecue (0.213), cake (0.413), and popular shop (0.203) respectively, account for the major portion of final intra-domain interest. Finally,  the top-100 candidate reviews contain all ground-truth items at the right of the figure, which are highly relevant to the above three interest capsules and confirm the reasonability and interpretability of M2GNN.

\section{Related work}
Our work is highly related to two subareas of recommendation, which are content-based and GNN-based CDR.

\noindent\textbf{Content-based CDR.}
The underlying assumption of this method is that the interests extracted from descriptive information in the content can be shared in different domains because of the consistency of user preferences. According to the method for modeling text features, there are two kinds of content-based CDR. 
The first way is directly projecting documents into the common low-dimensional space \cite{elkahky2015multi,fu2019deeply,hu2018mtnet}. 
RC-DFM \cite{fu2019deeply} learns latent vectors of reviews by applying a max pooling operation over pretrained word embeddings and utilizes the output of hidden layer under SDAE \cite{vincent2010stacked} framework to recommend items.
The second category of solutions fulfills high-level semantics extraction, also known as aspects \cite{song2017based,zhao2020catn}. 
CATN \cite{zhao2020catn} leverages the aspect gate as a soft on-off switch to control which latent feature derived from the text convolution layer is relevant to the aspect, and then performs pair-wise aspect matchings as the final rating prediction. 
Despite their success in processing text content, solutions in both categories don't specially take into consideration the corpus where the majority of tokens are noise. To our best knowledge, M2GNN is the first attempt to filter out noisy content and distill intra-domain interests simultaneously.

\noindent\textbf{GNN-based CDR.}
This category of algorithms aims to leverage the high-order connectivity of GNN by transforming domain-specific knowledge into graphs. There are two common graph data topologies: (1) user-item bipartite graph \cite{cui2020herograph,wang2017item,li2020heterogeneous,liu2020cross}.
BiTGCF \cite{liu2020cross} incorporates graph neural network to a bi-direction dual transfer learning approach, where each domain has a unique user-item bipartite graph and the output of each GNN layer is integrated across domains in the propagation process.
(2) heterogeneous graph  \cite{yin2019heterogenous,bi2020heterogeneous}.
HCDIR \cite{bi2020heterogeneous} transforms four types of insurance products with their properties and relations into an insurance graph and utilizes relational neighbor aggregation to learn embeddings in each domain, which is fed into an MLP to transfer knowledge.
However, all of the above models fail to incorporate review information into the graph explicitly and just use documents to generate initial representations. To our best knowledge, M2GNN is the first attempt to regard the words (tags) as one type of node in GNN-based CDR.

\section{conclusion}
In this paper, we aim to improve the review feed recommender system in the Dianping app. The proposed M2GNN adopts the GNN-based cross-domain framework to extract content-based interests via tags from other sites like search and consumption logs to enhance target recommendation. A key contribution is a novel two-step aggregation mechanism, which can distill intra-domain interests from a large number of tags, most of which are noises. Both offline and online experiments demonstrate the effectiveness and explainability of M2GNN.
For future work, we plan to investigate how to filter out unuseful tags via node sampling technology in GNN. 

\bibliographystyle{ACM-Reference-Format}
\bibliography{ref1}

\appendix

\end{sloppypar}
\end{document}